\documentclass{emulateapj}

\usepackage{amsmath}
\usepackage{amssymb}
\usepackage{natbib}
\usepackage{xspace}
\usepackage{graphicx}
\usepackage{rotate} 
\usepackage{subfigure}
\usepackage{float}
\usepackage[normalem]{ulem}

\shorttitle{A0535+26 Towards Quiescence}
\shortauthors{Rothschild et al.}

\begin{document}

\title{Observations of {\bf The} High Mass X-ray Binary A\,0535+26 in Quiescence} 

\author{Richard~Rothschild\altaffilmark{1}, Alex~Markowitz\altaffilmark{1,4,5}, Paul Hemphill\altaffilmark{1}, Isabel~Caballero\altaffilmark{2}, Katja~Pottschmidt\altaffilmark{3}, Matthias K\"uhnel\altaffilmark{4}, J\"orn~Wilms\altaffilmark{4}, Felix F\"urst\altaffilmark{6}, Victor Doroshenko\altaffilmark{7}, Ascension Camero-Arranz\altaffilmark{8}}

\affiliation{ 1 University of California, San Diego, Center for
  Astrophysics and Space Sciences, 9500 Gilman Dr., La Jolla, CA
  92093-0424, USA\\
 2 CEA Saclay, DSM/IRFU/SAp -UMR AIM (7158) CNRS/CEA/Universit\'e P. Diderot, Orme des Merisiers, Bat. 709, 91191 Gif-sur-Yvette, France\\
3 CRESST, UMBC, and NASA GSFC, Code 661, Greenbelt, MD 20771, USA\\
4 Dr. Karl-Remeis-Sternwarte and ECAP, Sternwartstr. 7, 96049 Bamberg, Germany\\
5 Alexander von Humboldt Fellow\\
6 Space Radiation Lab, MC 290-17 Cahill, California Institute of Technology, 1200 E. California Blvd, Pasadena, CA 91125\\
7 Institut f\"ur Astronomie und Astrophysik, Universit\"at T\"ubingen, Sand 1, 72076 T\"ubingen, Germany\\
8 Institut de Ci\`{e}ncies de l'Espai, (IEEC-CSIC), Campus UAB, Fac. de Ci\`{e}ncies, Torre C5, parell, 2a planta, 08193 Barcelona, Spain}
\email{rrothschild@ucsd.edu}


\begin{abstract}

We have analyzed 3 observations of the High Mass X-ray Binary A\,0535+26 performed by the \textsl{Rossi X-ray Timing Explorer}
(\textsl{RXTE}) 3, 5, and 6 months after the last outburst in 2011 February. We detect pulsations only in the second observation. 
The 3--20 keV spectra can be fit equally well with either an absorbed power law or absorbed thermal bremsstrahlung model.
Re-analysis of 2 earlier \textsl{RXTE} observations made 4 years after the 1994 outburst, original \textsl{BeppoSAX} observations 
2 years later, re-analysis of 4 \textsl{EXOSAT} observations made 2 years after the last 1984 outburst, and a recent 
\textsl{XMM-Newton} observation in 2012 reveal a stacked, quiescent flux level decreasing from $\sim$2 to 
$<1\times 10^{-11}$ ergs cm$^{-2}$ s$^{-1}$ over 6.5 years after outburst. Detection of pulsations during
half of the quiescent observations would imply that accretion onto the magnetic poles of the neutron star
continues despite the fact that the circumstellar disk may no longer be present. 
The accretion could come from material built-up at the corotation radius or from an isotropic stellar wind.

\end{abstract}



\section{Introduction}

A\,0535+26 is a Be/X-ray binary system, discovered by \textsl{Ariel V} during a giant outburst in 1975 \citep{Rosenberg75}. 
The binary system consists of the neutron star pulsar A0535+26 and O9.7-B0 IIIe optical companion
HDE\,245770 \citep{Steele98}. The neutron star moves in an eccentric orbit with \textsl{e} = 0.47, an orbital period 
$P_\mathrm{orb}$ = 111.07$\pm$0.07 days, and exhibiting a pulse period of $\sim$103.25 s 
\citep[][and references therein]{Camero-Arranz12}. 
The estimated distance to the system \textsl{d} = 2 kpc \citep{Steele98}. Extensive reviews of the
system are given in \citet{Giovannelli92} and \citet{Caballero09}. 
A\,0535$+$26 belongs to a  class of high mass systems, called Be/X-ray binaries, known for having outbursts where matter accretes onto the magnetic poles 
via an accretion disk that is filled at periastron 
passage from a circumstellar decretion disk of the primary. When this stellar disk retreats, the 
accretion disk shrinks and the source enters 
quiescence. This allows the observer to study other source(s) of accretion that may be present. \citet{Campana02} have
studied high mass systems in quiescence where the neutron star spin period is relatively short ($P_\mathrm{spin} <$ 5 s), 
and they conclude that X-ray emission is due to material falling onto the magnetospheric boundary or burning on the neutron 
star surface. Longer period high mass systems are less well studied.

\begin{figure*}
 \includegraphics[width=4.0in]{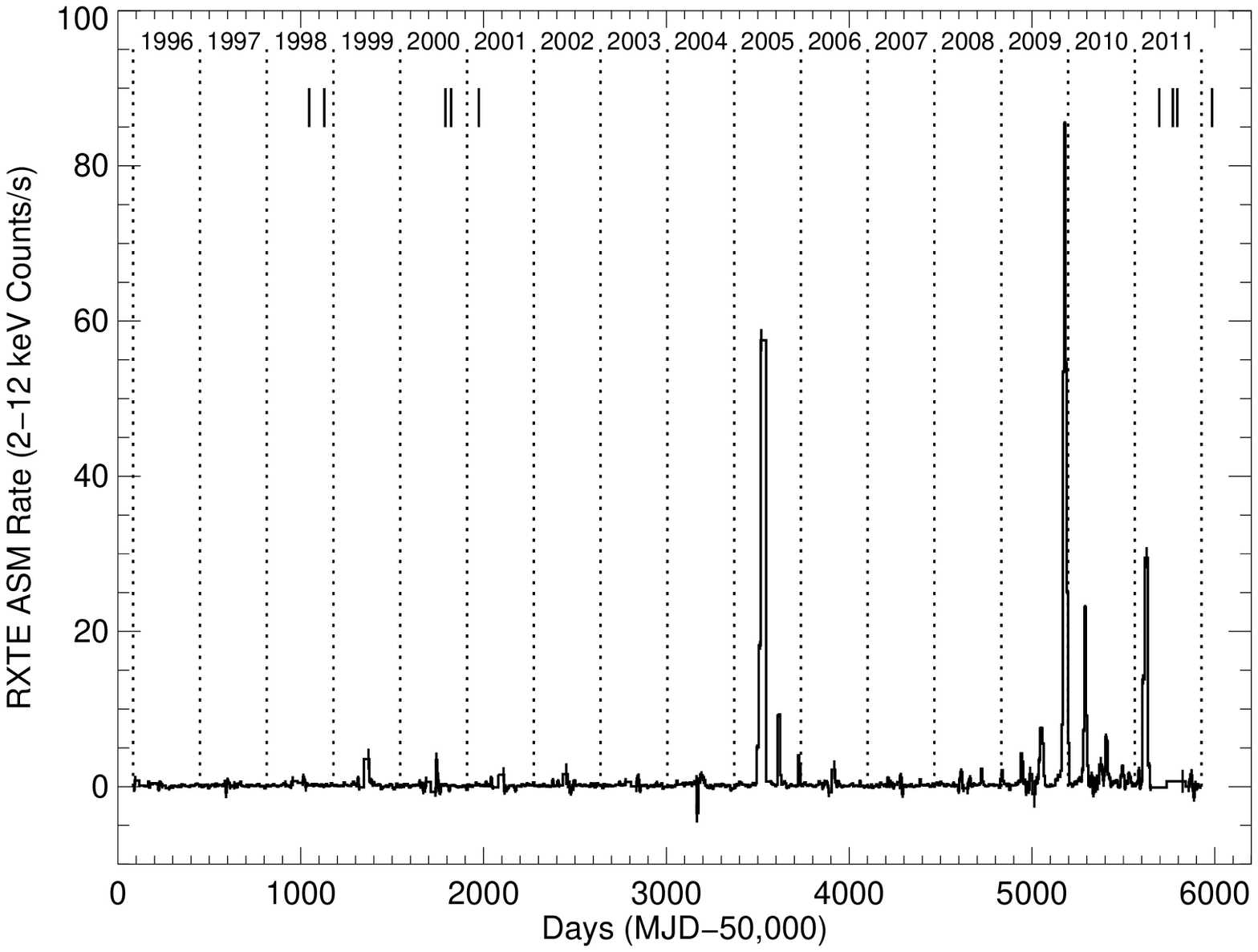}
 \includegraphics[bb=31 26 469 378,width=3.5in]{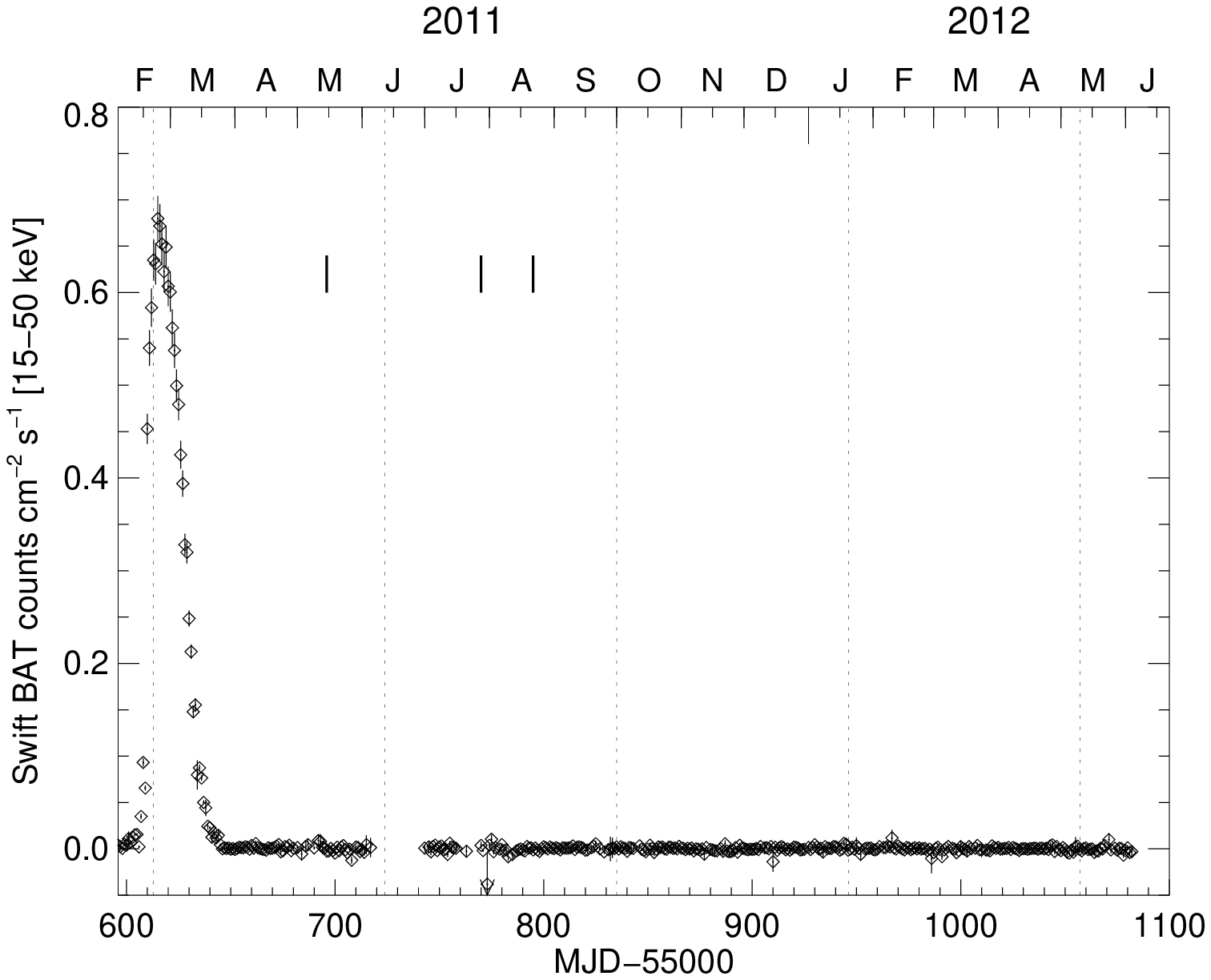}\\
  \caption{(Left) \textsl{RXTE}/ASM 2--12 keV weekly light curve for the transient A\,0535$+$26 throughout the 
  \textsl{RXTE} mission and (Right) the \textsl{Swift}/BAT 15--50 keV daily light curve from 2011 to present (bottom). 
  Yearly boundaries are shown by vertical dotted lines in the plot \textbf{on the left}, and periastron passages are marked 
  by vertical dotted lines in the plot on the right. The dates of the \textsl{RXTE, XMM-Newton},
  and \textsl{BeppoSAX} observations are noted by short vertical lines in the plot on the left, and the three new \textsl{RXTE} 
  pointed observations are marked on the plot on the right. See Table~\ref{tab:quiescence} for dates and observatories.
  No outbursts have been seen since the 2011 February outburst.}
  \label{fig:bat}
\end{figure*}

The X-ray intensity of A\,0535+26 varies by almost three
orders of magnitude with three basic intensity states: 1) quiescence, 
2) normal, or type I, outbursts, generally associated with periastron passages, and 3) giant, or type II, outbursts 
that may occur at any orbital phase. The companion, HDE\,245770, has an equatorial circumstellar 
disk whose size has varied over time. It is this material that 
drives the normal outbursts at periastron. 
The giant outbursts may arise from large asynchronous mass outflows from the companion.
An indicator of the disk size is the H$\alpha$ equivalent width, as well as that of He I and the 
overall visual magnitude \citep{Grundstrom07}.
The results of monitoring of these quantities are given in Figure 1 of \citet{Camero-Arranz12} for the past 37 years.
A large reduction in the H$\alpha$ strength occurred in 1998 (MJD $\sim$51000) in conjunction with the ceasing of outbursts
for 7 years. New outbursts began anew with the giant outburst of 2005 (Fig.~\ref{fig:bat}~(Left)).
Beginning in 2009, the H$\alpha$ strength again declined with an apparent leveling-off at the time of the 
2011 February outburst. Since then the H$\alpha$ strength has continued to decline (Camero-Arranz pvt. comm.) 
and  no outbursts have been detected.

Since its discovery, nine giant outbursts have been detected: 
in 1977 December \citep{Giovannelli92}, 1980 October  \citep{Nagase82}, 1983 June  \citep{Sembay90}, 
1989 March/April  \citep{Makino89}, 1994 February  \citep{Finger94},
2005 May/June  \citep{Tueller05}, 2009 December \citep{Wilson-Hodge09}, 2010 October \citep{Mihara10}, and 
2011 February \citep{Camero-Arranz12}. Normal outbursts are seen at periastron passages \citep[see Figure 4 in][]{Motch91},
but may not occur for years at a time.
After the 1998 normal outburst, the source went into the quiescent state until resuming in 2005 (Fig.~\ref{fig:bat}~(Left)).  
The last outburst was the giant one in 2011 February. 
Since then, A0525$+$26 has exhibited no outbursts at all (Fig.~\ref{fig:bat}~(Right)).

\begin{figure*}
  \includegraphics[width=3.0in]{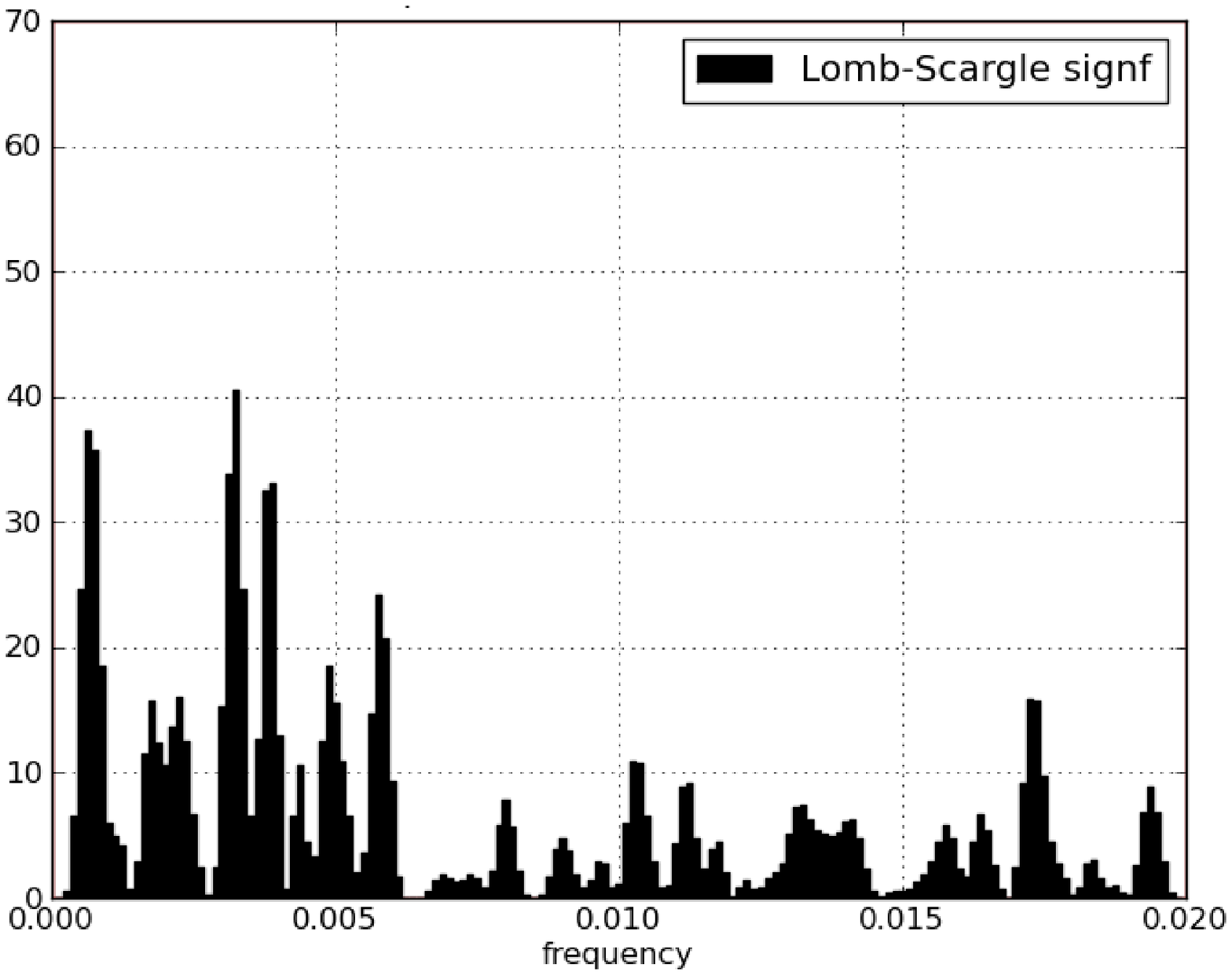}
  \includegraphics[width=3.2in]{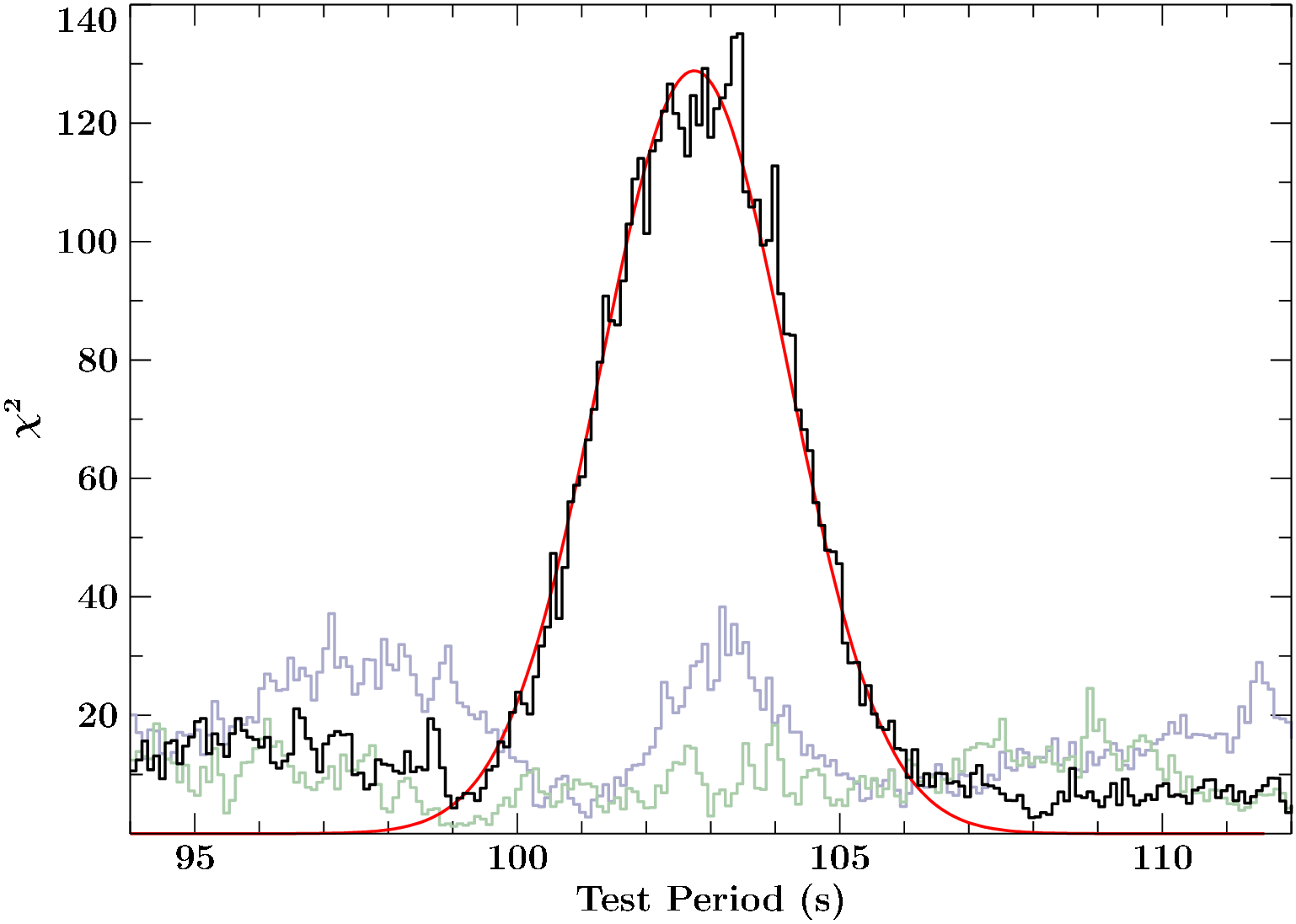}\\
  \includegraphics[width=3.0in]{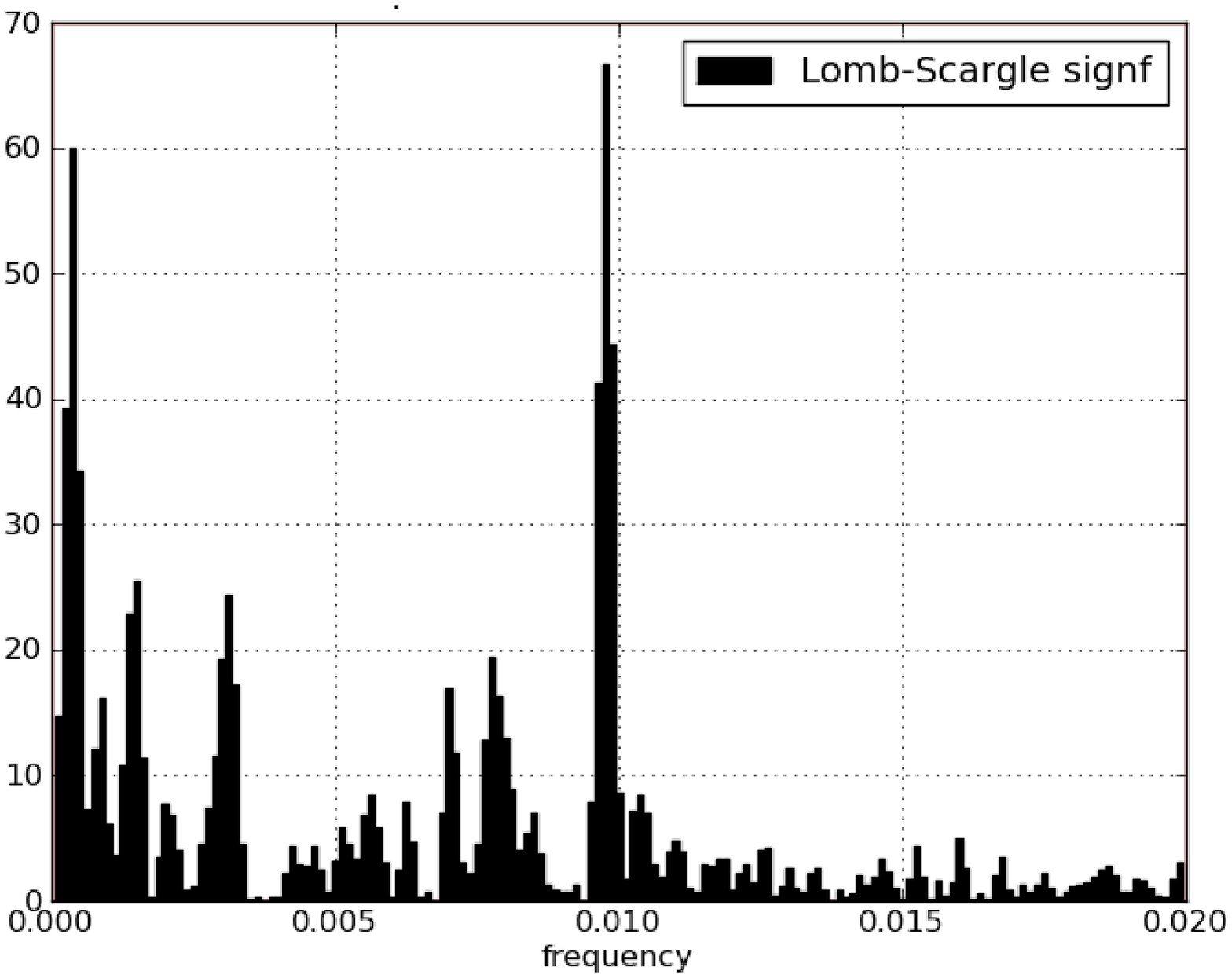}
  \includegraphics[width=3.2in]{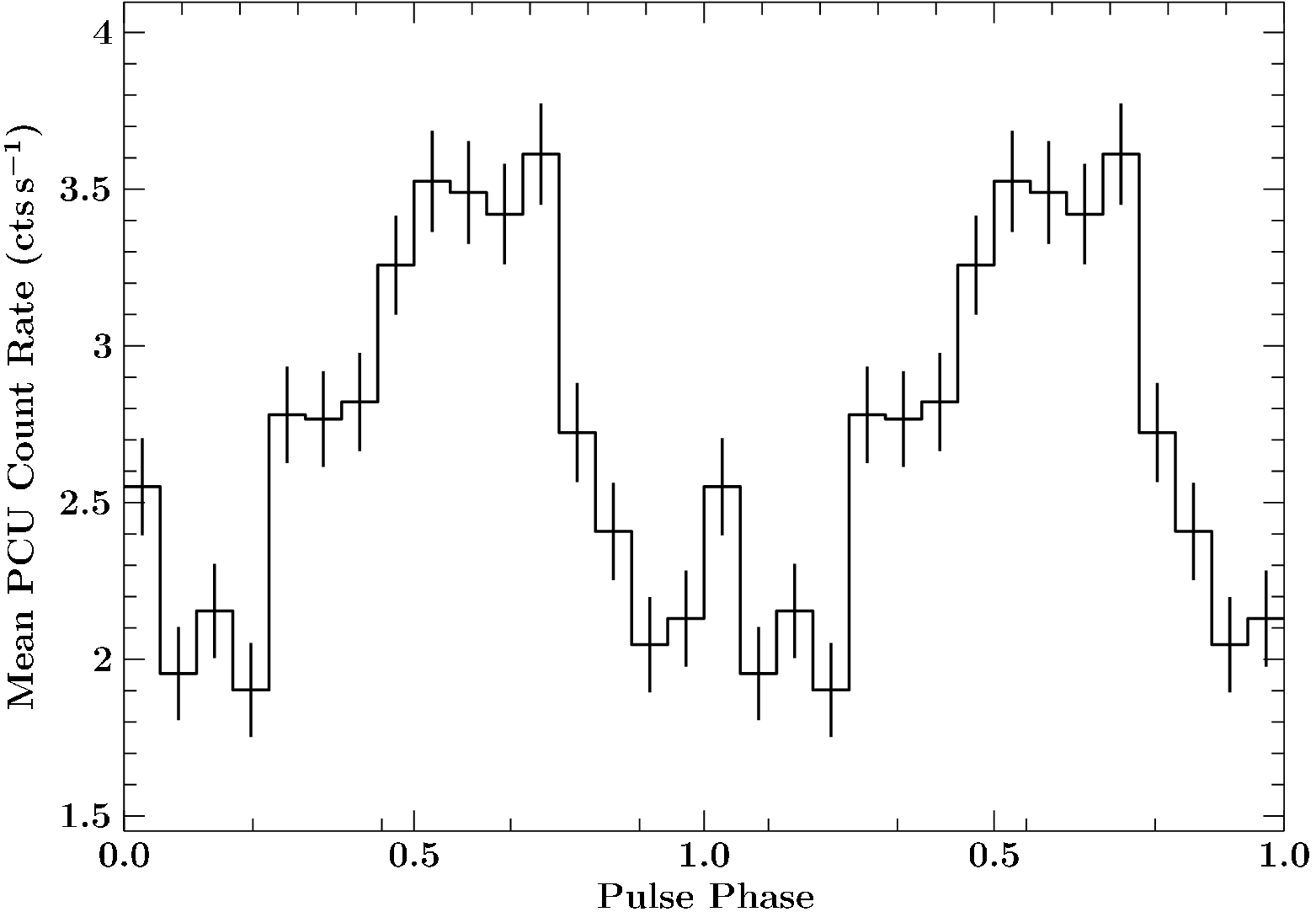}\\
  \includegraphics[width=3.0in]{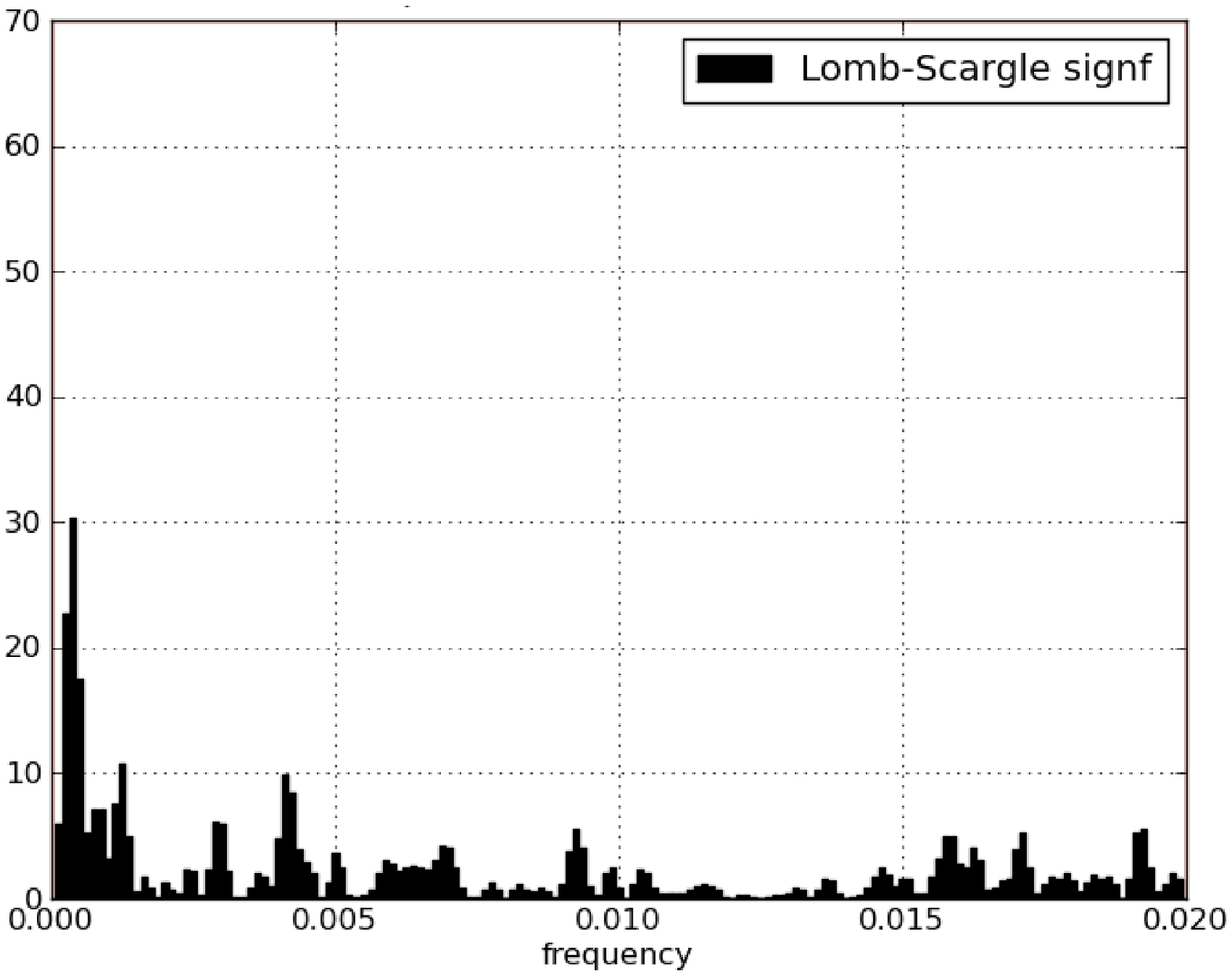}\\
  
   \caption{(Left column, top to bottom)  Lomb-Scargle significance plots for the 2011 May, 2011 July, 
   and 2011 August observations, respectively.
   Only the 2011 July data show highly significant detection of pulsations near 103 s (0.0097 Hz). The significance at 
   very low frequency is due to variations with periods on the order of the observation time.
   (Top-Right) $\chi^2$-distributions resulting of epoch folding of observations 2011 May 15 (blue), 
   2011 July 28 (black) and 2011 August 22 (green). The red line shows the best Gaussian fit to the second observation.
   (Middle-Right) Pulse profile of the 2011 July 28 observation using the peak period of 103.41(28) s.  
   The profile is repeated once to enhance visibility.}
  \label{fig:epochfold}
\end{figure*}

In quiescence, the system is expected to be in the centrifugally inhibited regime \citep{Illarionov75}, 
preventing the continual accretion of matter onto the neutron star. Nevertheless, all
historic observation campaigns during quiescence have found pulsations in at least part of their
data, indicating that matter is still being accreted along the magnetic field lines. 
Observations of the quiescent state have been performed by \textsl{EXOSAT, RXTE, BeppoSAX}, and \textsl{XMM-Newton}.
The source was observed four times by \textsl{EXOSAT} between 1985 and 1986 
and pulsations were detected during two of the observations \citep{Motch91}. 
\textsl{RXTE} performed two observations in quiescence between
1998 August and 1998 November \citep{Negueruela00}, and pulsations were detected during the 
1998 November observation, while in August only a weak 
indication ($< 3\sigma$) was found for periodicity at the nominal pulse period. 
\textsl{BeppoSAX} observed the source
during quiescence in 2000 September-October and 2001 March \citep{Orlandini04}. 
\citet{Mukherjee05} divided the \textsl{BeppoSAX} data into 5 ever-decreasing count-rate bins, and 
significantly detected pulsations in all but the faintest. 

We report here on three \textsl{RXTE} observations made in 2011 after the giant outburst of 2011 February, 
and pulsations were only detected in the second of the new observations. Combining these three observations
with the previous observations and a new \textsl{XMM-Newton} observation made a year after the last 
outburst (Doroshenko pvt. comm.), we present a detailed study of A0535$+$26 in quiescence, where accretion onto the magnetic poles continues, possibly from matter accumulated at the magnetospheric radius or a stellar wind.
Section 2 describes the data reduction and analysis, while Section 3 discusses the results and gives the 
conclusions of this investigation.

\section{Data Reduction and Analysis}

\begin{deluxetable}{rrrrr}
\tabletypesize{\scriptsize}
\tablecaption{Observations of A\,0535$+$26 with \textsl{RXTE}\label{tab:obs}}
\tablewidth{0pt}
\tablehead{
\colhead{Date} & \colhead{ObsID} & \colhead{Rate\tablenotemark{a,b}} & \colhead{Lvt\tablenotemark{c}} 
& {Phase\tablenotemark{d}}}
\startdata
2011 May 15 & 96421-01-02-00 & 3.9$\pm$0.1& 2,432 & 0.749   \\
2011 July 28 & 96421-01-03-00 & 2.3$\pm$0.1 & 2,944 & 0.415  \\
2011 August 22 & 96421-01-04-00 & 1.4$\pm$0.1 & 2,672 & 0.640 
\enddata
\tablenotetext{a}{PCU2 3--90 keV background subtracted counting rate in counts/s}
\tablenotetext{b}{Uncertainties in rates are 68\% confidence}
\tablenotetext{c}{The livetime in seconds}
\tablenotetext{d}{Based upon $P_\mathrm{orb}$=111.1 days, MJD(Periastron)=53613.0}
\end{deluxetable}

\vspace*{0.05in}

\textsl{RXTE} observed A\,0535$+$26 in quiescence on three occasions in 2011 in order to study its spectral and 
temporal behavior (Table~\ref{tab:obs}).  The data reduction was carried out using HEASOFT version 6.11 software, and XSPEC 
version 12.7.0 \citep{Arnaud96}. The data were extracted observing the standard UCSD/IAAT/ECAP filtering criteria of 
pointing within 0.01 degrees of the target, elevation above the Earth's horizon greater than 10$^\circ$, time since the 
center of the last South Atlantic Anomaly passage of 30 minutes, and an electron rate of less than 0.1. 

\begin{deluxetable}{rccccc}
\tabletypesize{\scriptsize}
\tablecaption{Quiescient Observations of A\,0535+26\label{tab:quiescence}}
\tablewidth{3.0in}
\tablehead{
 \colhead{Date} & \colhead{Satellite} & \colhead{MJD} 
& \colhead{$\Delta$T\tablenotemark{a}} & \colhead{Phase\tablenotemark{b}} & \colhead{Pulses?}
}
\startdata
2011 May 15          & \textsl{RXTE} & 55696 & 77 & 0.749 & No\\
2011 July 28          & \textsl{RXTE} & 55770 & 155 & 0.415 & Yes\\
2011 Aug 22     & \textsl{RXTE} & 55795 & 180 & 0.640 & No\\
2012 Feb 28 & \textsl{XMM}  & 55985 & 370 & 0.350 & Yes\\
\hline
1985 Sept 23 & \textsl{EXOSAT} & 46331 & 600 & 0.545 & Yes\\
1986 Feb 10     & \textsl{EXOSAT} & 46471 & 740 & 0.284 & No\\
1986 Feb 18     & \textsl{EXOSAT} & 46479 & 748 & 0.212 & Yes\\
1986 Mar 9            & \textsl{EXOSAT} & 46498 & 767 & 0.041 & No\\
\hline
1998 Aug 21            & \textsl{RXTE} & 51046 & 1455 & 0.105 & Yes\\
1998 Nov 12    & \textsl{RXTE} & 51129 & 1538 & 0.358 & Yes\\
2000 Sept 4-5 & \textsl{BeppoSAX} & 51791& 2200 & 0.400 & No\\
2000 Oct 5-6       & \textsl{BeppoSAX} & 51822 & 2231 & 0.121 & No\\
2001 Mar 6-7          & \textsl{BeppoSAX} & 51974 & 2383 & 0.753 & Yes
\enddata
\tablenotetext{a}{Days since the peak of the previous outburst; 2011 February 23, 1994, August 27, or 1984 February 1}
\tablenotetext{b}{Based upon $P_\mathrm{orb}$=111.1 days, MJD(Periastron)=53613.0}
\end{deluxetable}
\vspace*{0.1in}

We have also re-analyzed the 1998 \textsl{RXTE} observations of A\,0535$+$26 reported by \citet{Negueruela00} 
and the 1985--1986 \textsl{EXOSAT} observations of \citet{Motch91} using updated response and background files\footnote{Our re-analysis appears to show that \citet{Motch91} 
have a typographical error in their Table 2 --- the 1--10 keV flux should be $\times$10$^{-11}$ ergs cm$^{-2}$ s$^{-1}$.}.
The re-analysis of the \textsl{EXOSAT} spectra for A\,0535$+$26 was performed by downloading the spectral and 
response files for the \textsl{EXOSAT} ME detector from the HEASARC for the 4 observations. The \textsl{XMM-Newton} results will
be the subject of a separate publication, and only the 2--10 keV flux is presented here. The \textsl{BeppoSAX} results
are from the publications of \citet{Orlandini04} and \citet{Mukherjee05}. In addition, we have used the 
abundances of \citet{Wilms00} and cross sections of \citet{Verner96}.  Table~\ref{tab:quiescence}
summarizes the 13 observations made in quiescence since 1984 in order of their time since the last outburst preceding them.

\subsection{Timing}

To perform a period search, Proportional Counter Unit 2 \citep[PCU2;][]{Jahoda06}  lightcurves from the observations 
in 2011 July (ObsId 96421-01-03-00) and 
2011 August (ObsId 96421-01-04-00) were extracted in 
data mode B\_250ms\_128M and the lightcurve from the observation  in 2011 May (ObsId 96421-01-02-00) in GoodXenon mode. 
Due to the low luminosity of the source, no significant flux was detected above $\sim$20 keV, and therefore only 
data from pulse height channels 
5 to 49, corresponding to energies between $\sim$3 and $\sim$20\,keV, were used in the timing analysis 
(See Table~\ref{tab:obs} for the comparative counting rates). 
For each observation the data were collected into consecutive 2 s time bins. With uniformly sampled and uninterrupted
data, the Lomb-Scargle algorithm was applied over the frequency range of 0 to 0.020 Hz (periods greater than 50 s) 
in 160 frequency steps (approximately 1 s steps in period). The resulting Lomb-Scargle significance is plotted 
versus frequency for the three data sets
in Fig~\ref{fig:epochfold} (Left column). The probability of a frequency component 
occurring in pure Gaussian noise is Pr(occ) $\cong$ (0.01)$^{s/4.6}$, where $s$ is the significance. 
Thus a significance of 66.7, for the middle observation at 0.0097 Hz, implies a very significant detection. 
The interpretation of the actual significance is limited by systematic effects, such as aliasing and non-Gaussian variations. 
The peak of the Lomb-Scargle plot for the middle observation was at 102.36 s. Note that the Lomb-Scargle 
analysis assumes sinusoidal variations and Gaussian noise, while the pulse profile of A\,0535$+$26 is somewhat 
sinusoidal (Fig.~\ref{fig:epochfold} (Lower-Right)) and the statistics are Poisson. Epoch folding for periods near the 
Lomb-Scargle peak value was used to further define the pulse period.
The first and last observations produce very little Lomb-Scargle significance near 0.0097 Hz (103 s), and we conclude 
that pulsing was not detected for those observations. Epoch folding for those data was performed to validate the 
Lomb-Scargle analysis.

An iterative epoch folding method was utilized whereby pulse-to-pulse 
changes in luminosity were taken into account by averaging the count rate over the pulse period and subtracting the
resulting pulse averaged lightcurve from the original one. This increases the strength of a signal in the epoch 
folding $\chi^2$-distribution \citep{Larsson96}. Periodic signals were searched between 94 and 112 s, with 200 
test periods in between and 8 phase bins. The resulting $\chi^2$ distribution for the 2011 July observation is shown 
in Figure~\ref{fig:epochfold}~(Top-Right), and the corresponding 16 bin folded light curve for a peak period of 
103.48(28)s is given on the Middle-Right. If one compares the
second observation profile to those seen in outburst \citep[e.g. ][]{Caballero07}, this 3--20 keV profile resembles the higher 
energy ($>$15 keV) ones in outburst with the broader minimum, as compared to lower energy ones with a narrow 
feature defining flux minimum. It also resembles many of the previous quiescent 2--10 keV profiles. 
The uncertainty in the epoch fold period for the 2011 July data is at the $1\sigma$-confidence limit and was 
determined by Monte Carlo simulations.  The 2011 July pulse period is consistent with 
the detected pulse period of $\sim$103.25\,s \citep{Camero-Arranz12} during the  last outburst in February 2011, 
and considering that the pulse period changes relatively little during non-outburst times. 
It is surprising that the pulsations were only detected in the 2011 July observation, which has an intermediate 
flux of the three observations. Even if the lack of a detection of pulsations in the last of the three observations 
could be attributed to insufficient statistics, the non-detection in the brightest observation remains a puzzle.
With exposure roughly equivalent to the 2011 July observation and higher counting rate, one would expect it to have 
been more sensitive to pulsations. The energy spectra of all 3 observations are quite similar 
(Fig.~\ref{fig:spectral_power}), so the presence 
or lack of pulsations does not appear attributable to some kind of state change. 

\subsection{Spectra}

\begin{figure*}
\plottwo{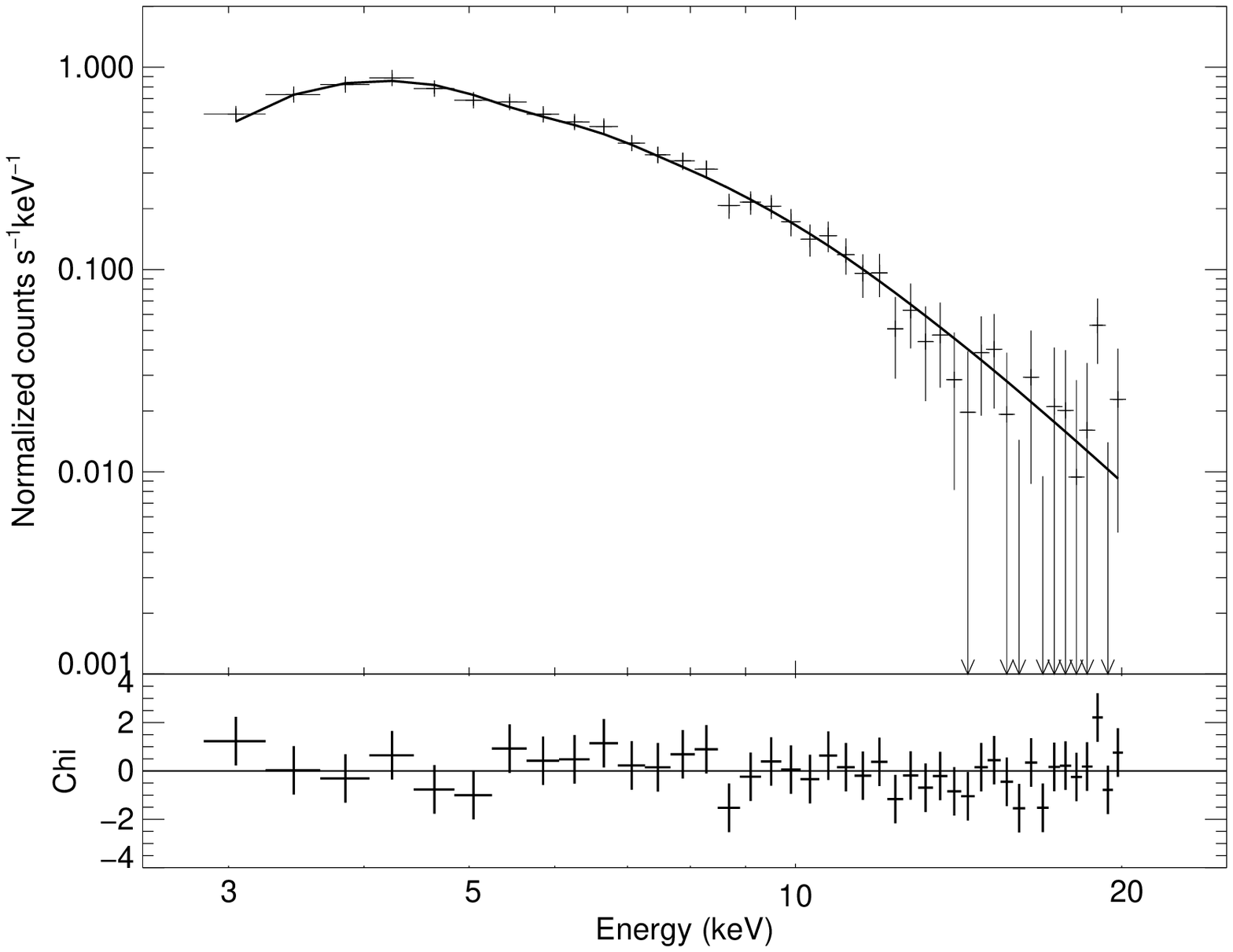}{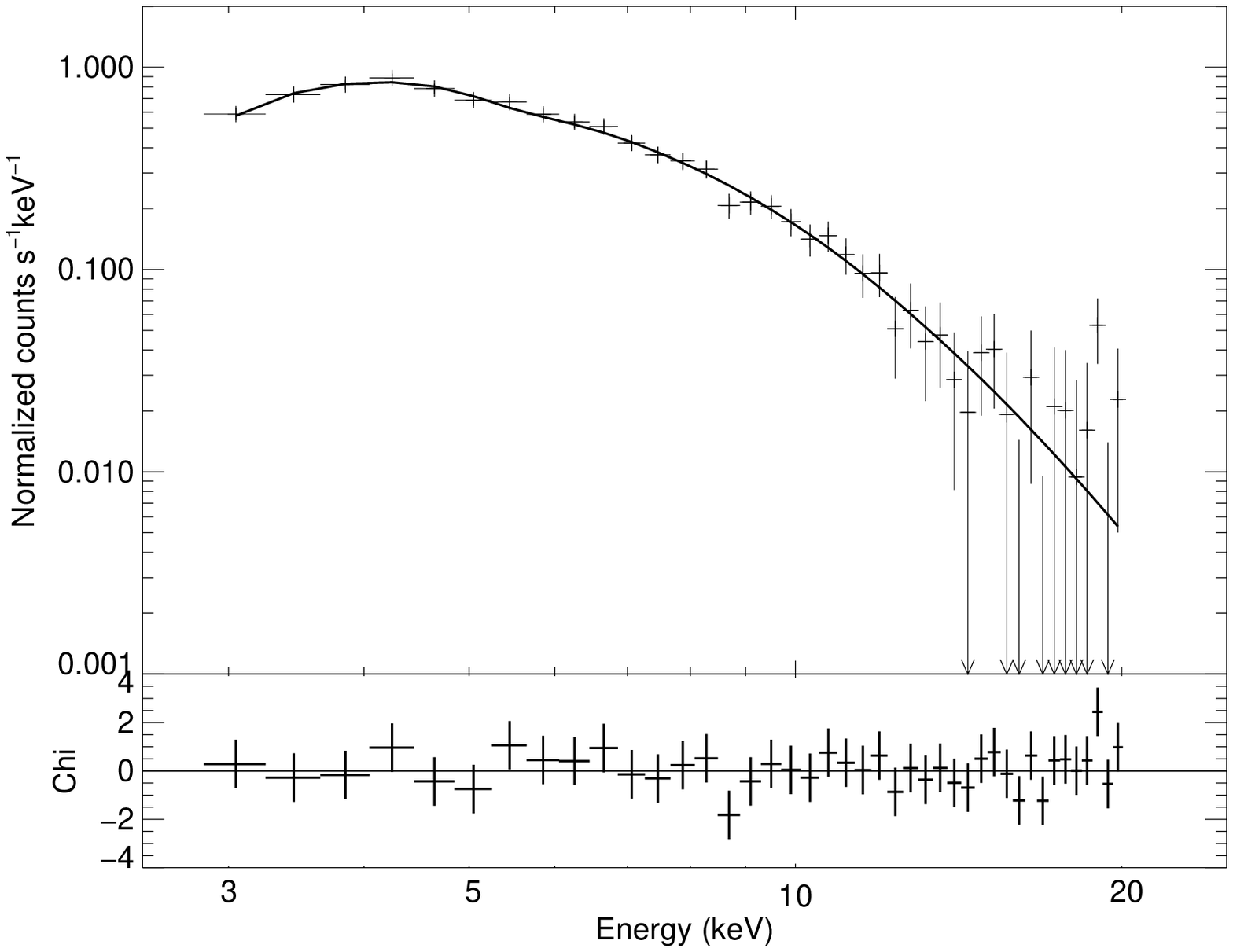}\\
\plottwo{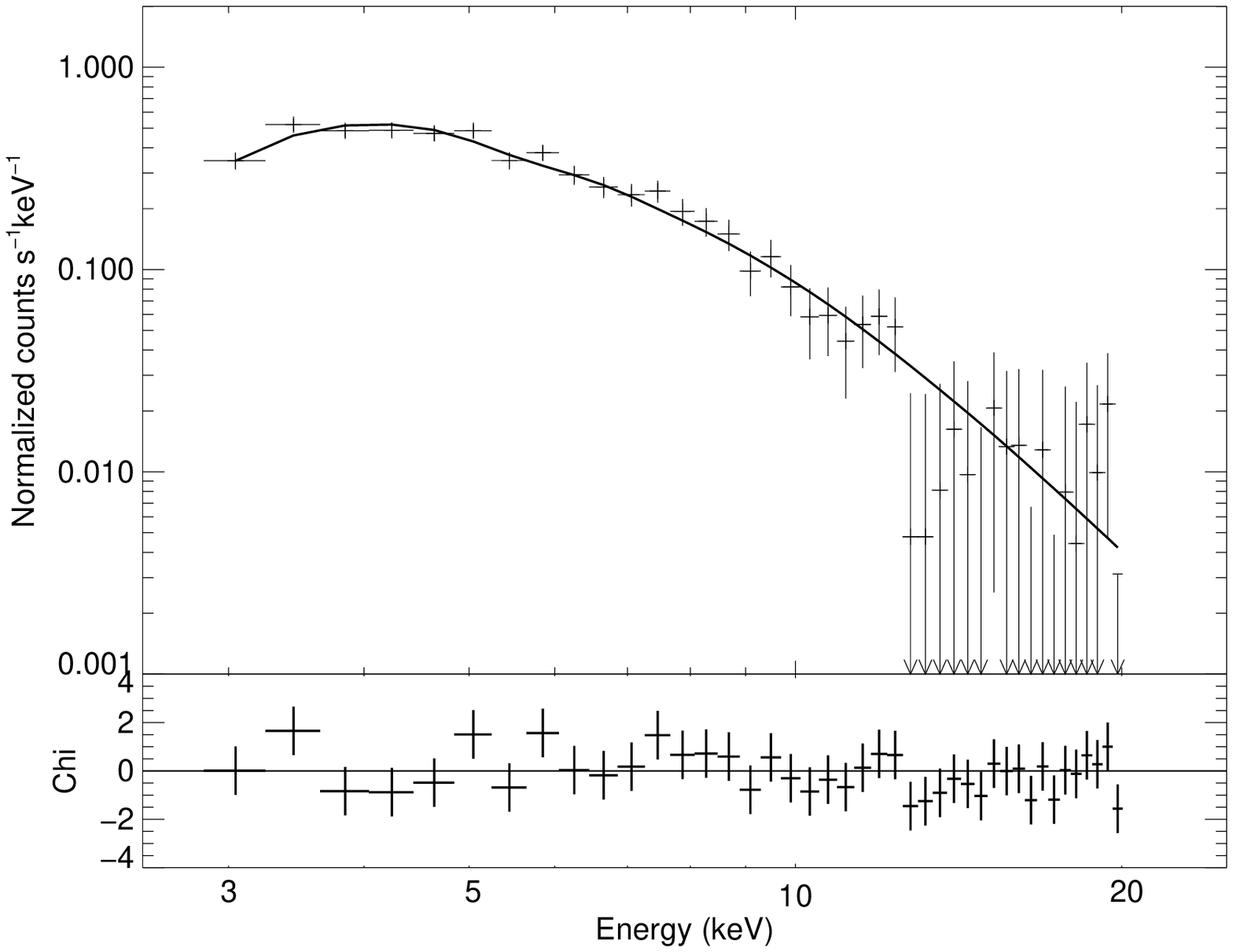}{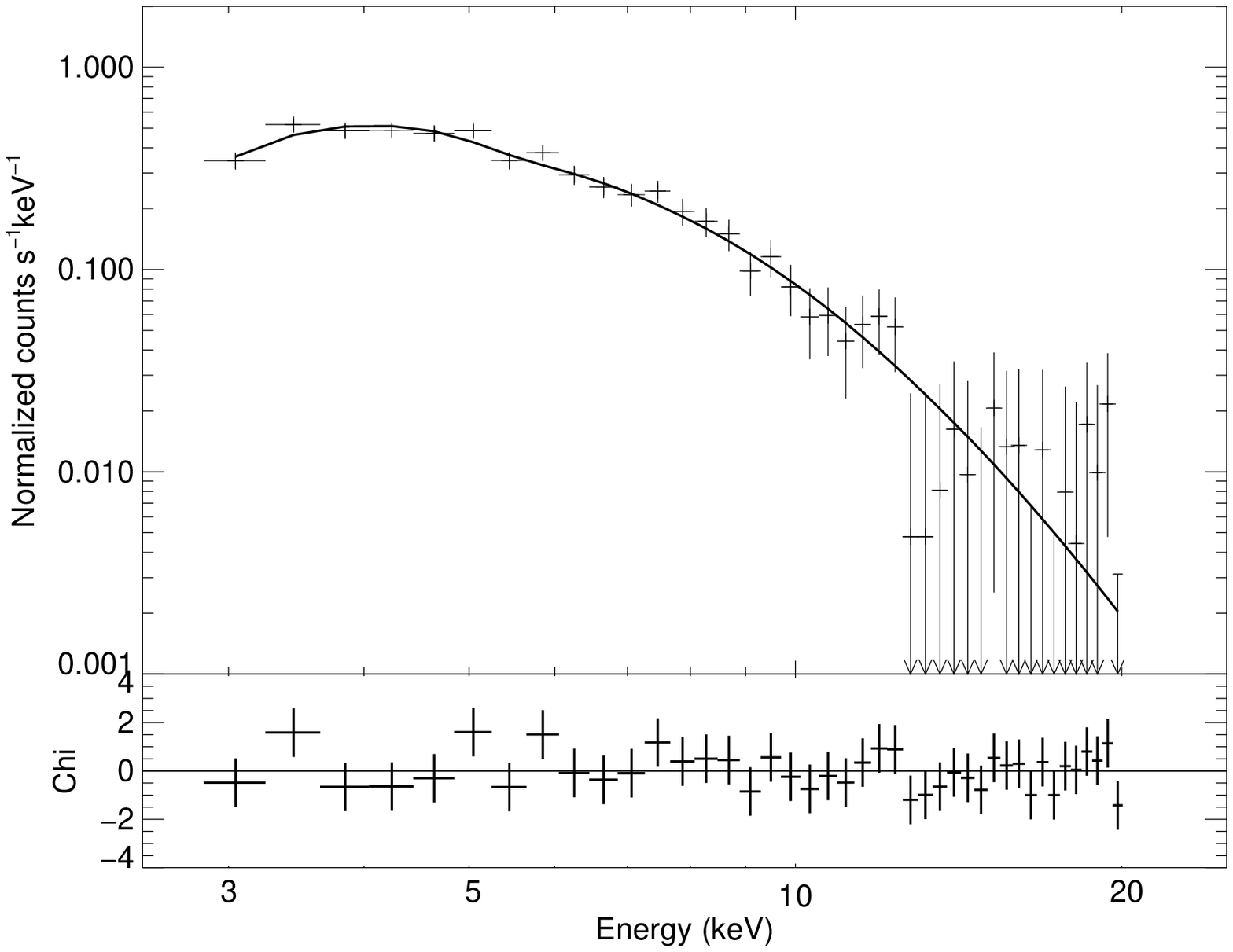}\\
\plottwo{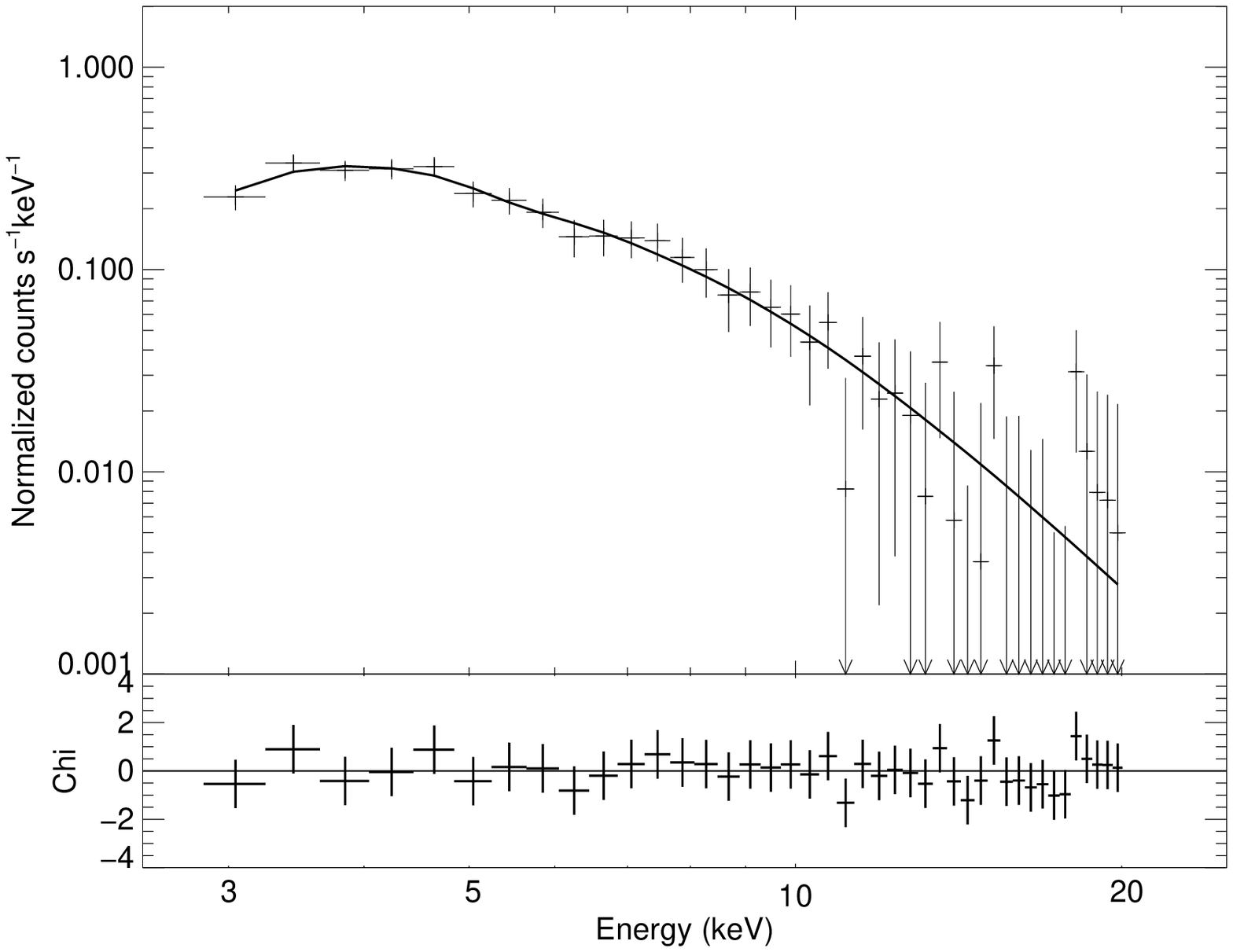}{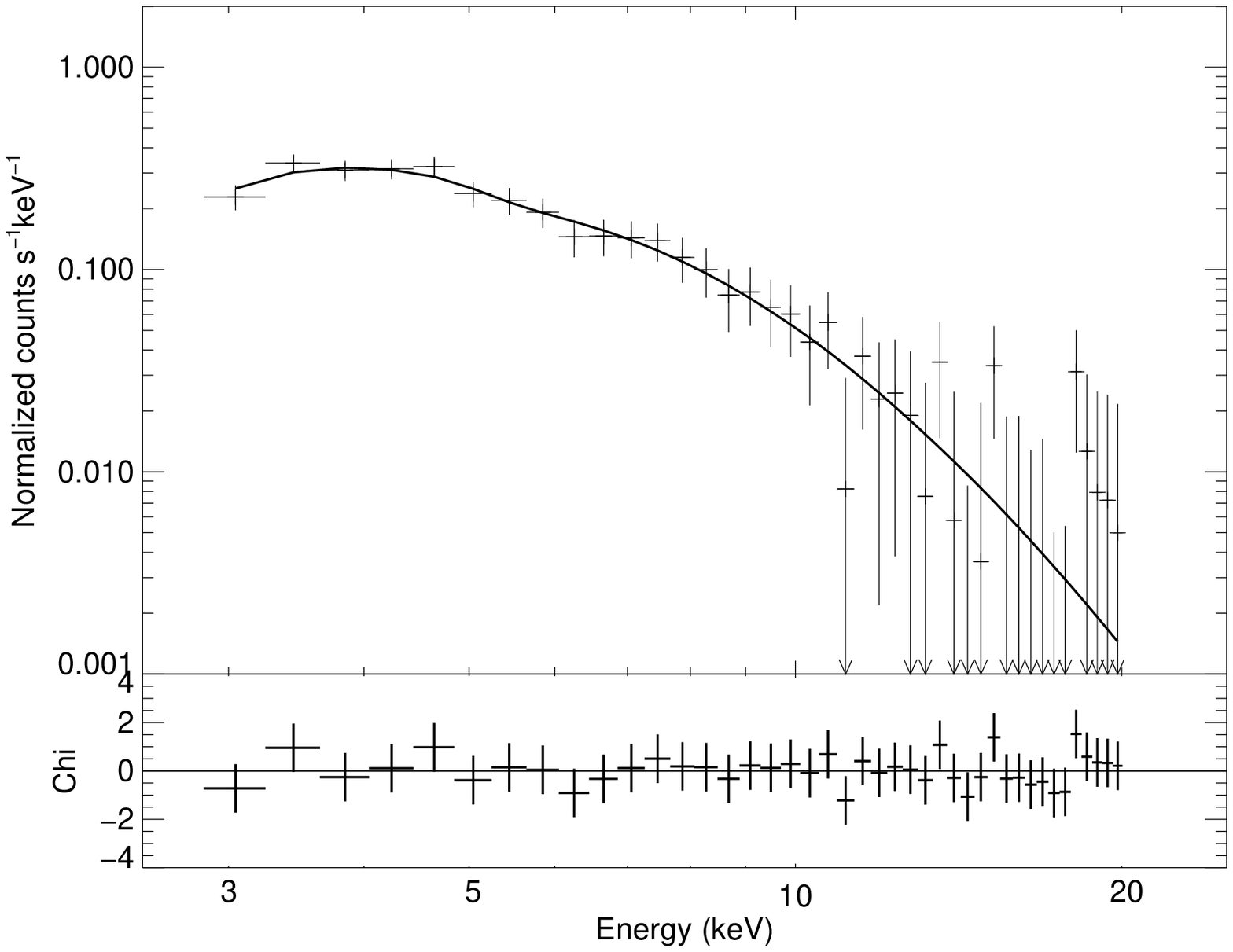}\\
\caption{Power law (left) and thermal bremsstrahlung (right) spectral fits to \textsl{RXTE}/PCA spectra for 
the three epochs of observing 
(2011 May 15, July 28, and August 22, respectively). The lower panels in each show the number of standard deviations of the 
model from the data.}
\label{fig:spectral_power}
\end{figure*}

The spectra for each \textsl{RXTE} observation were generated from the PCU2 Standard2 Data files, which represent the full 
(2--90 keV) energy range in 129 variable width spectral bins. These histograms are independent of the individualized data 
modes for a given observation. The PCU2 Faint background estimate, provided by the PCA team and based 
upon detector rates acquired while observing blank fields during the mission,
was used in fitting of background subtracted data. The normalization of the background histogram 
depends upon the dead time estimate, which is also based upon 
average detector counting rates. The XSPEC recorn function 
was initially used to correct for the small difference between the estimated and actual PCU2 background normalization, 
which cannot be measured directly due to the Proportional Counter Array detectors being non-imaging. 
Consequently the background must be modeled. 
Comparison of results with and without the recorn function in the model showed that these data were insensitive 
to this background optimization, and thus results are given for fits without the recorn function included in the modeling.
The two spectral models fitted to the three observations in the 3--20 keV range were an absorbed power 
law (\textsc{cflux*(tbabs*pegpwrlw)}) and absorbed thermal bremsstrahlung (\textsc{cflux*(tbabs*bremss)}), 
both modified by a maximum in Galactic line of sight absorption \citep[fixed at 4.5$\times 10^{21}$ cm$^{-2}$,][]{Kalberla05}. 
\textsc{cflux} was the basis for the measurement of 2--10 keV flux plus 90\% errors. 
The simple power law was used since the statistics did not warrant the use of a more complex model with a 
high energy cutoff, as utilized for fitting outburst spectra. Both models gave 
acceptable and similar fits (Fig.~\ref{fig:spectral_power}, Tables~\ref{tab:power} and~\ref{tab:bremss}). 
The errors reported in the two Tables are 90\% uncertainties.
The power law and thermal bremsstrahlung best-fit values for absorption at the source differ by about a factor of two, 
with the power law one averaging about 9$\times$10$^{22}$ cm$^{-2}$ and bremsstrahlung one averaging a bit less than 
4$\times$10$^{22}$ cm$^{-2}$. This is due to the curvature of the bremsstrahlung spectrum with respect to that 
of a power law at low energies.  Above $\sim$10 keV, the bremsstrahlung spectrum rolls over with respect to the 
power law, but poor statistics at these energies preclude the determination of which model provides the better representation of the 
incident spectrum.

The 1998 \textsl{RXTE} spectral data for A\,0535$+$26 were extracted with the same acceptance criteria as the 
2011 \textsl{RXTE} data, and re-analyzed with updated response and background files.
For the present \textsl{RXTE} re-analysis, only data from PCU2 were used for consistency with 
the later \textsl{RXTE} results. Only 2$\sigma$ upper limits to the column
density were found, and the best-fit values of the power law indices were less than but consistent with those found by 
\citet{Negueruela00}. When applying the bremsstrahlung model, the column density upper limits were again lower than 
those for the power law fits, and the electron temperatures were somewhat higher, but again consistent with the
values found by \citet{Negueruela00}. Tables~\ref{tab:power} and \ref{tab:bremss} give the best-fit values from the
re-analysis. While \citet{Motch91} reported average \textsl{EXOSAT} 
values for the best-fit parameters for both power law and thermal bremsstrahlung models, we report the 
individual observation values here (Tables~\ref{tab:power} and \ref{tab:bremss}). We found an average power law 
index of 2.27 and an average electron temperature of 7.5 keV. No simultaneous fitting between different instruments 
was performed due to the variations in column density, flux, and power law index. Consequently the relative 
normalizations between the instruments are assumed to be unity.

\begin{deluxetable}{rcccc}
\tabletypesize{\scriptsize}
\tablecaption{Power Law Spectral {\bf Fit} Results\label{tab:power}}
\tablewidth{0pt}
\tablehead{
\colhead{Date} & \colhead{$N_\mathrm{H}$\tablenotemark{a}} & \colhead{Photon Index} & \colhead{Flux\tablenotemark{b}} 
& \colhead{$\chi^2_\nu$\tablenotemark{c}}
}
\startdata
2011 May 15 & 9.1$^{+1.4}_{-1.2}$ & 2.40$^{+0.11}_{-0.07}$ & 4.01$^{+0.23}_{-0.11}$ & 0.67\\
2011 July 28 & 9.1$^{+4.1}_{-2.4}$ & 2.59$^{+0.24}_{-0.13}$ & 2.37$^{+0.18}_{-0.12}$ & 0.75\\
2012 Aug 22 & 5.0$^{+6.1}_{-4.0}$ & 2.44$^{+0.43}_{-0.28}$ & 1.55$^{+0.18}_{-0.16}$ & 0.42 \\
\tableline
1998 Aug 21\tablenotemark{d} & $<$11.5 & 2.30$^{+0.53}_{-0.60}$ & 0.79$^{+0.15}_{-0.10}$ & 0.50\\
1998 Nov 12\tablenotemark{d} & $<$13.7 & 2.40$^{+0.63}_{-0.59}$ & 0.96$^{+0.21}_{-0.10}$ & 0.46\\
\tableline
1985 Sept 23\tablenotemark{e} & 2.4$^{+0.5}_{-0.5}$ & 1.83$^{+0.13}_{-0.07}$ & 2.46$^{+0.09}_{-0.13}$ & 1.33\\
1986 Feb 10\tablenotemark{e} & 4.0$^{+2.4}_{-1.5}$ & 2.42$^{+0.50}_{-0.34}$ & 1.18$^{+0.11}_{-0.11}$ & 0.96\\
1986 Feb 18\tablenotemark{e} & 2.5$^{+1.2}_{-1.0}$ & 1.98$^{+0.27}_{-0.27}$ & 1.44$^{+0.11}_{-0.11}$ & 0.81\\
1986 Mar 9\tablenotemark{e} & 6.7$^{+3.8}_{-4.4}$ & 2.85$^{+0.73}_{-1.00}$ & 1.56$^{+0.34}_{-0.16}$ & 0.89
\enddata
\tablenotetext{a}{Column density in units of 10$^{22}$ cm$^{-2}$}
\tablenotetext{b}{Absorbed 2--10 keV power law flux in units of 10$^{-11}$ ergs cm$^{-2}$ s$^{-1}$}
\tablenotetext{c}{Chi-square per degree of freedom for 39 degrees of freedom for \textsl{RXTE} and 52 degrees of freedom for \textsl{EXOSAT}}
\tablenotetext{d}{Re-analysis of \citet{Negueruela00} \textsl{RXTE} observations}
\tablenotetext{e}{Re-analysis of \citet{Motch91} \textsl{EXOSAT} observations}
\end{deluxetable}
\vspace*{0.1in}

\begin{deluxetable}{rcrrr}
\tabletypesize{\scriptsize}
\tablecaption{Bremsstrahlung Spectral {\bf Fit} Results\label{tab:bremss}}
\tablewidth{0pt}
\tablehead{
\colhead{Date} & \colhead{$N_\mathrm{H}$\tablenotemark{a}} & \colhead{$kT$\tablenotemark{b}} & \colhead{Flux\tablenotemark{c}} 
& \colhead{$\chi^2_\nu$\tablenotemark{d}}
}
\startdata
2011 May 15 & 3.7$^{+1.8}_{-0.8}$ & 8.22$^{+0.88}_{-0.95}$ & 4.22$^{+0.19}_{-0.17}$ & 0.60\\
2011 July 28  & 3.7$^{+3.2}_{-1.6}$ & 6.59$^{+0.97}_{-1.05}$ & 2.48$^{+0.19}_{-0.14}$ & 0.65\\
2012 Aug 22  & $<$5.4                  & 7.44$^{+2.30}_{-2.15}$ & 1.62$^{+0.14}_{-0.17}$ & 0.41 \\
\tableline
1998 Aug 21\tablenotemark{e} & $<$7.0                  & 9.06$^{+8.13}_{-4.04}$ & 0.80$^{+0.13}_{-0.09}$ & 0.48\\
1998 Nov 12\tablenotemark{e} & $<$8.1            & 7.17$^{+6.62}_{-2.64}$ & 1.00$^{+0.16}_{-0.11}$ & 0.42\\
\tableline
1985 Sept 23\tablenotemark{f} & 1.6$^{+0.8}_{-0.4}$ & 11.14$^{+2.15}_{-2.19}$ & 2.42$^{+0.10}_{-0.12}$ & 1.25\\
1986 Feb 10\tablenotemark{f} & 2.4$^{+1.9}_{-1.1}$ & 4.66$^{+3.33}_{-1.15}$ & 1.11$^{+0.17}_{-0.08}$ & 0.93\\
1986 Feb 18\tablenotemark{f} & 1.6$^{+1.1}_{-0.6}$ & 8.42$^{+6.81}_{-2.53}$ & 1.40$^{+0.14}_{-0.10}$ & 0.81\\
1986 Mar 9\tablenotemark{f} & 4.1$^{+3.4}_{-2.8}$ & 3.84$^{+8.26}_{-1.43}$ & 1.51$^{+0.38}_{-0.16}$ & 0.90
\enddata
\tablenotetext{a}{Column density in units of 10$^{22}$ cm$^{-2}$}
\tablenotetext{b}{Electron temperature in keV}
\tablenotetext{c}{Absorbed 2--10 keV flux $\times 10^{-11}$ ergs cm$^{-2}$ s$^{-1}$}
\tablenotetext{d}{Chi-square per degree of freedom for 40 degrees of freedom for \textsl{RXTE} and 52 degrees of freedom for \textsl{EXOSAT}}
\tablenotetext{e}{Re-analysis of \citet{Negueruela00} observations}
\tablenotetext{f}{Re-analysis of \citet{Motch91} observations}
\end{deluxetable}

\subsection{Correlations and Trends}

\begin{figure*}
 \plottwo{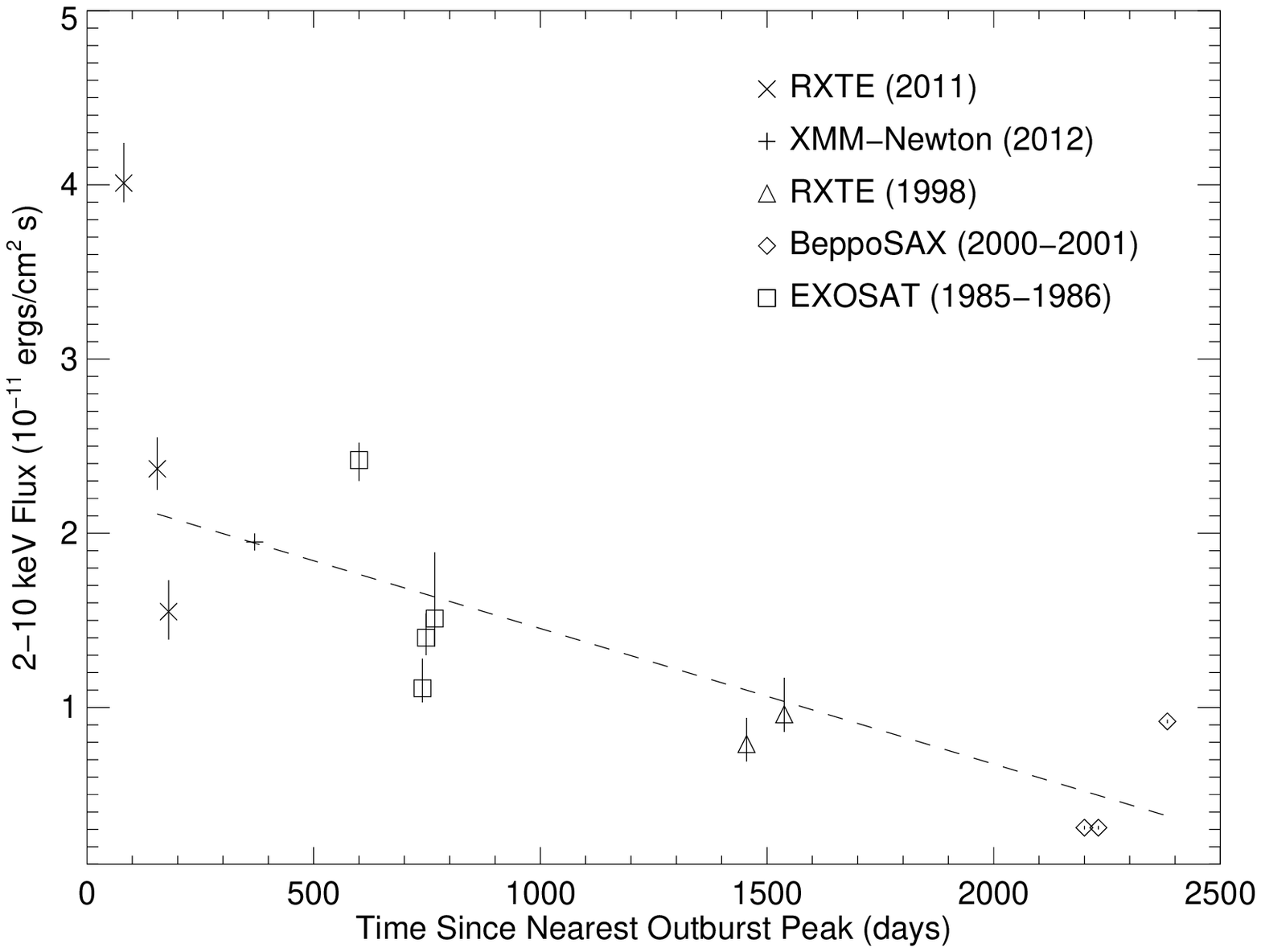}{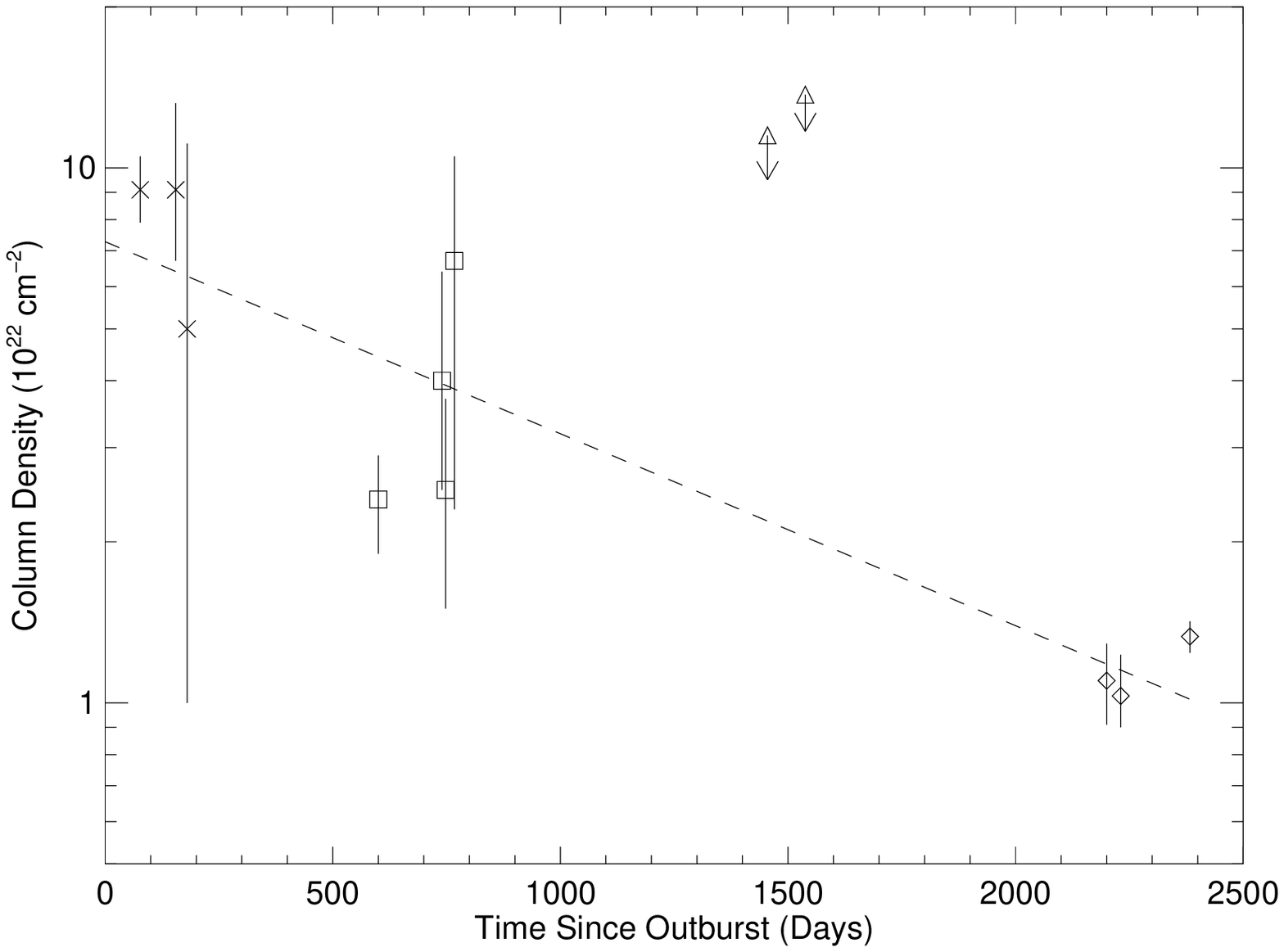}\\
\caption{(Left) Plot of 2--10 keV fluxes observed by \textsl{RXTE, XMM-Newton, EXOSAT}, and \textsl{BeppoSAX} 
as a function of time since the peak of the previous outburst. The X's represent the present \textsl{RXTE} fluxes, the 
plus sign is the \textsl{XMM-Newton} flux, the triangles are the reprocessed \textsl{RXTE} fluxes, the squares are 
the reprocessed \textsl{EXOSAT}values, and the diamonds are the 
\textsl{BeppoSAX} fluxes. The dashed line is a least squares fit to all but the first \textsl{RXTE} flux, and indicates 
the declining trend in the flux.
(Right) Plot of the column density N$_\mathrm{H}$ versus time since outburst with the same symbols. The dashed line represents a least-squares fit to all the data. The \textsl{RXTE} and \textsl{EXOSAT} points are from the reanalysis reported here, and the \textsl{BeppoSAX} points are from \citet{Orlandini04}. 
\label{fig:time_since}}
\end{figure*}

The present 2011 \textsl{RXTE} observations that occurred 3, 5, and 6 months after 
the last outburst (2011 Feb 23) show a clear steeply dropping flux (Fig.~\ref{fig:time_since}). However, when the 
\textsl{XMM-Newton} value from 12 months after the outburst is added, the interpretation is less clear, and a flux
 level of $\sim$2$\times10^{-11}$ ergs cm$^{-2}$ s$^{-1}$ 6--12 months after the outburst is indicated.
The addition of the earlier 
\textsl{RXTE, BeppoSAX} and \textsl{EXOSAT} observations performed 2--6 years into quiescence sample the declining flux from 
0.25 to 6 years after an outburst (Fig.~\ref{fig:time_since}~(Left)), albeit not the same outbursts and not all either normal 
or giant outbursts. The dashed line in Fig.~\ref{fig:time_since}~(Left)
is a least squares fit to all but the first \textsl{RXTE} point, and reflects a Pearson correlation coefficient of $-$0.785
(a null hypothesis probability of 0.012).
If taken at face value, the quiescent  flux values indicate a slowly declining quiescent level with a slope 
of 3.8$\times10^{-12}$ ergs cm$^{-2}$ s$^{-1}$ per day after these 3 outbursts. The 2011--2012 \textsl{RXTE} data  
show a slightly raised flux in the first observation, as do the \textsl{EXOSAT} data. This may be indicative of small levels of flaring
during quiescence, which would be yet another indication of continuing, variable accretion. It may also be an effect of orbital 
phase in an isotropic stellar wind.

\begin{figure*}
 \plottwo{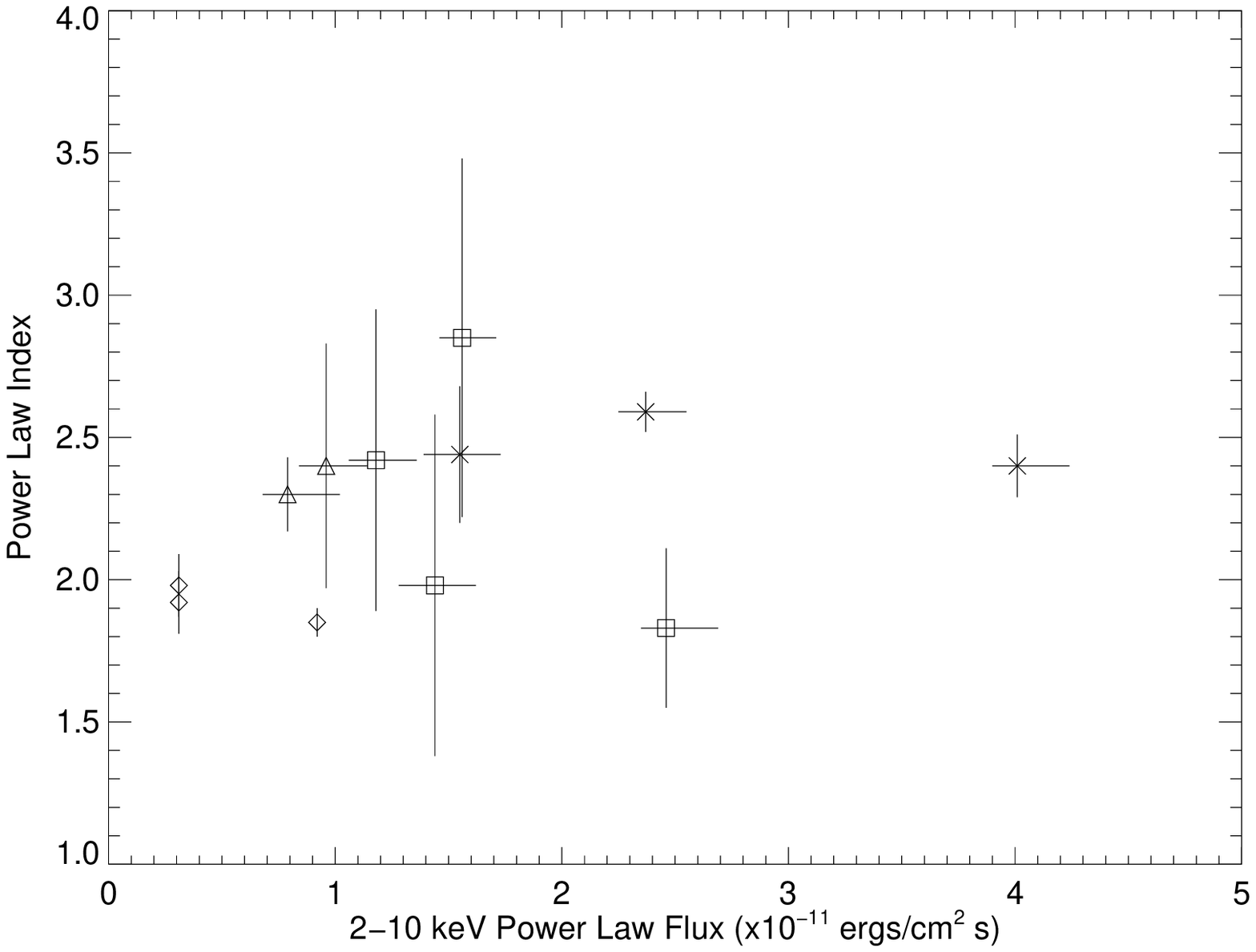}{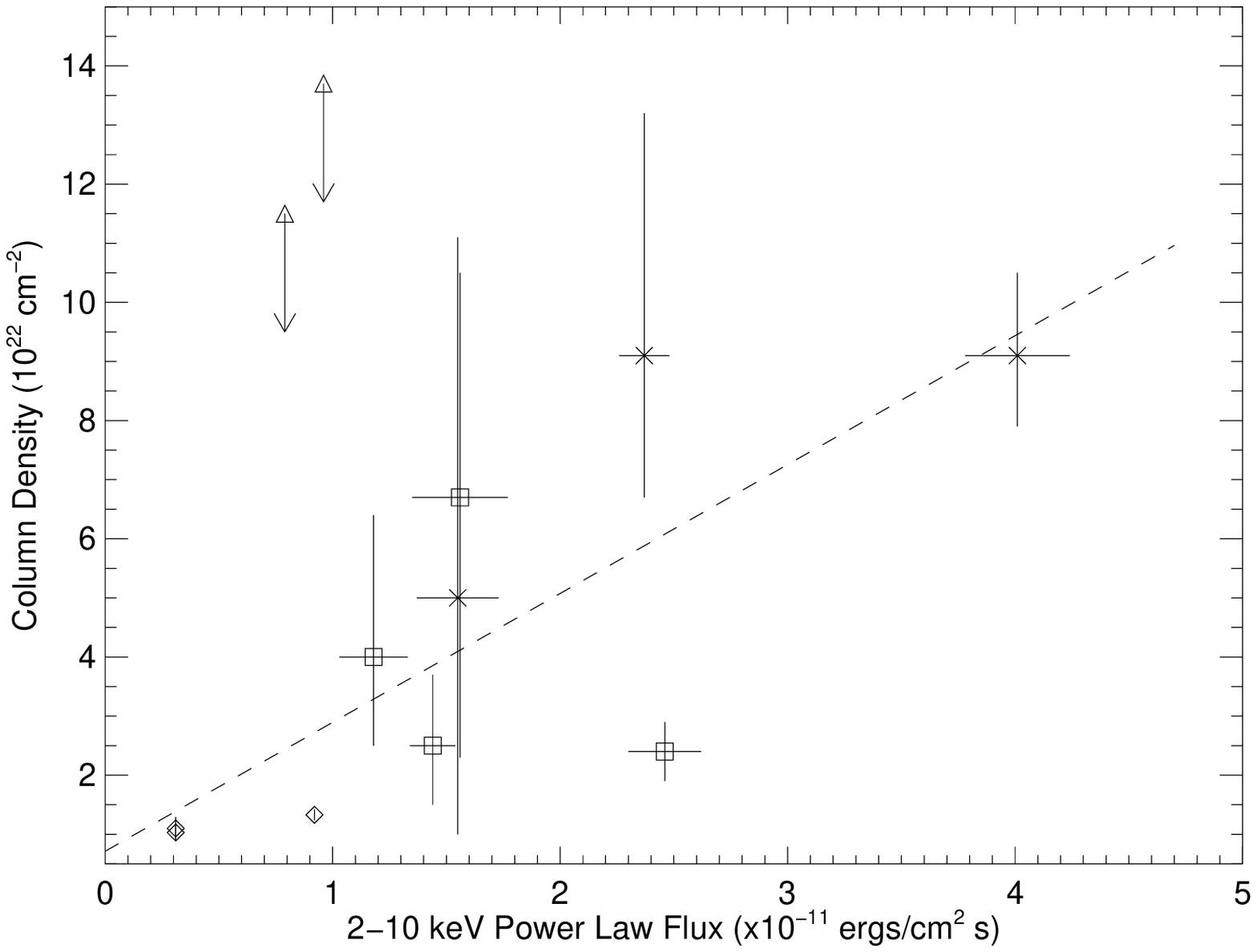}\\
 \caption{(Left) Plot of the power law index versus 2--10 keV power law flux. Note that no correlation is evident. X's represent the new \textsl{RXTE} data, triangles the re-analyzed \textsl{RXTE} data, and squares the reanalyzed \textsl{EXOSAT} data. The diamonds represent the \textsl{BeppoSAX} data \citep{Orlandini04}. (Right) Plot of the column density versus 2--10 keV power law flux with the same symbols. The dashed line represents the least-squares fit to the data.
 \label{fig:index_flux}}
 \end{figure*}

As the observed intensity decreased linearly as a function of time since outburst, the column density decreased 
exponentially (Fig.~\ref{fig:time_since}~(Right)). 
The Pearson correlation coefficient for the log of $N_\mathrm{H}$ versus time since outburst is $-$0.897 (a null hypothesis probability of 0.001).  
This naturally results in column density increasing with the 2--10 keV power
 law flux (Fig.~\ref{fig:index_flux}~(Right)), and in this case the Pearson correlation coefficient is 0.772 (a null hypothesis probability of 0.015). Finally, the power law index does
 not appear to correlate with the power law flux (Fig.~\ref{fig:index_flux}~(Left)). 

\section{Discussion and Conclusions}

A\,0535$+$26 continues to show pulsations, as seen in most but not all of the observations. The fact that 
the first and brightest of the 3 new \textsl{RXTE} observations in quiescence did not exhibit detectable pulsations, 
while the second, $\sim$50\% weaker observation did with a similar live time, is strong evidence that the 
pulsations are not always present in quiescence. This would indicate in general that accretion onto the polar 
caps was still occurring, but that the accretion may not always be channeled to the magnetic poles. 
In quiescence the neutron star in A\,0535$+$26 is expected to be centrifugally inhibited \citep{Campana02}, and
plasma entry into the magnetosphere may be mediated by various instabilities affecting the extent of the emission region \citep{Arons76}.

\citet{Okazaki01} have modeled the outbursts from Be/X-ray binaries using a viscous decretion model, where the
presence of outbursts depends upon the neutron star orbit with respect to various resonances of the gas in the system. 
If the neutron star's orbit is outside a given resonance, matter from that resonance will not flow onto the neutron star,
whereas if it is within a given resonance, the matter will flow. \citet{Okazaki01} state for A\,0535$+$26 that the disk 
is truncated at the 4:1 resonance when outbursts are detected, and that a slight decrease in viscosity or disk 
temperature would lower the disk radius to the 5:1 resonance and thus have the system go into dormancy. 
The observation that the quiescent luminosity is declining in the composite light curve (Fig.~\ref{fig:time_since}) 
could imply that an accumulated mass at the corotation radius \citep{Haigh04} is being depleted over time. Then quiescence 
in A\,0535$+$26 would be characterized by reduced accretion from the magnetospheric boundary, 
while the bright outbursts would be characterized by high levels of accretion from the companion's circumstellar disk.

Another possibility is that the companion may have a stellar wind in addition to that from the circumstellar disk. This would 
mean that the system could transition from disk accretion to stellar wind accretion. If the disk were to recede, as 
the H$\alpha$ observations imply in 
1998 and 2011--2012 (Figure 1 of \citet{Camero-Arranz12} and V. McBride, pvt. comm.), one might expect the system to transition
to spherical wind accretion dominating over that of the disk. If  there is a sufficient stellar wind to provide 
wind accretion as in GX 301-2 \citep[e.g. ][]{Suchy12}, then the generation of X-rays and pulsing seen in quiescence might be explained. The very small number of quiescent data points after each outburst as a function of orbital phase precludes any test of variations with orbital phase, which might have been evidence of stellar wind accretion. 

\citet{deLoore84} adopt a wind mass loss rate for HDE\,245770 of 10$^{-8}$ M$_\odot$ yr$^{-1}$ from measurements 
of UV resonance lines and infrared fluxes. Spherical Bondi accretion for a 1.4 M$_\odot$ neutron star at a distance of 
267 light-seconds ($a \sin i$ for the A\,0535$+$26 system) and a 2000 km s$^{-1}$ wind velocity at infinity 
\citep[e.g., Table 1 in ][]{Kudritzki00} would predict the mass accretion rate onto the neutron star to be $\sim 3 \times 10^{-5}$ 
of the wind mass loss rate. If the bolometric quiescent X-ray luminosity is about twice the 2-10 keV luminosity, 
or ~2.5$\times 10^{34}$ ergs s$^{-1}$ for a distance of 2 kpc, then one would predict a wind mass loss rate 
of 10$^{-9}$ M$_\odot$ yr$^{-1}$, if the conversion efficiency of gravitational to X-ray energy is 0.1. 
This estimate is 10\% of the \citet{deLoore84} value, which may be reasonable if 90\% of the wind mass 
loss is due to the circumstellar disk. \citet{Waters88} predict a factor of 10$^4$ for the difference in X-ray luminosities 
between polar and equatorial mass flows for HDE 245770 (10$^{31}$ ergs s$^{-1}$ vs 10$^{35}$ ergs s$^{-1}$). 
If the orbit of the neutron star
is inclined to the equator of the companion, and if the mass flow changes with angle from the pole, the neutron 
star may sample a mass flow that produces fluxes in the 10$^{34}$ ergs s$^{-1}$ range.

Should we see orbital variations of the flux due to the eccentric orbit in the \textsl{RXTE} data that was taken within 
one orbital period? A 2000 km s$^{-1}$ wind would predict at most a factor of ~$\sim$3 variation in flux between 
the 3 \textsl{RXTE} phases. The \textsl{RXTE} fluxes are in agreement with this limited variation. 

The spectra of A\,0535$+$26 \citep[Tables \ref{tab:power} and \ref{tab:bremss} plus the results from \textsl{BeppoSAX;}][]  
{Orlandini04,Mukherjee05} indicate a decreasing amount of absorption at the source as time since an outburst 
increases (Fig.~\ref{fig:time_since}~(Right)). When comparing the various values of $N_\mathrm{H}$  versus orbital phase 
(given in Table \ref{tab:quiescence}) it is clear that this material is present 
in the line of sight independent of orbital phase. Thus, it must be either all pervasive in the 
system (the wind?) or very close to the neutron star (a declining accumulation of matter at the corotation radius?). For the former, 
variations in $N_\mathrm{H}$ might imply a variable stellar wind density, which in turn would result in a correlated variable X-ray flux
 (Fig.~\ref{fig:index_flux}~(Right)). For the latter, the disk might 
thin as the material is drained from it. This could be further evidence for a continual draining of the material trapped 
near the corotation radius \citep{Dangelo12}. 

The power law index is independent of flux over the flux range of 1--4 $\times 10^{-11}$ ergs cm$^{-2}$ s$^{-1}$ 
(Fig.~\ref{fig:index_flux}). In addition, the power law indices measured for all of the quiescent observations are 
significantly steeper ($\sim$ 2.5 versus $<$ 1) than those seen during outbursts \citep[e.g. ][]{Caballero07}. 
A steeper power law would imply less Compton up-scattering of lower energy photons due to either a thinning of 
the hot electon scattering plasma, or a cooling of the electron distribution.
This may be further indication that the accretion mode is different between outburst and quiescence. 

The pulse profile exhibited by the 2011 July 28 \textsl{RXTE} data is single peaked and somewhat sinusoidal, as has been seen
in the past for observations made at quiescent luminosities, as noted above. \textsl{CGRO}/BATSE observations hint 
at a change in the pulse profile to the single peak variety that occurs below 10$^{37}$ ergs cm$^{-2}$ s$^{-1}$  \citep{Bildsten97}. 
This change from the double peaked structure of the outburst pulse profiles possibly indicates a change in the accretion regime.
For low luminosities the beam pattern is expected to be more pencil beam-like, and more like a fan beam at higher 
luminosities \citep[][and references therein]{Becker12}.  

\subsection{Conclusions}

\textsl{RXTE} observations of A\,0535$+$26 months after a giant outburst were combined with three other sets of 
multiple observations of A0535\,$+$26 from other X-ray observatories after other outbursts to reveal that 
they all had a similar (within a factor of 2), or slowly decreasing flux level. 
At least half of the observations detected pulsations, indicating that 
accretion onto the magnetic poles was still occurring, but perhaps intermittently. Optical observations of H$\alpha$ 
emission at the times of the 
measurements indicated that the 1998 and 2011 data were taken during times of a highly diminished Be star disk, allowing
for the possibility of sources of accreted material other than that responsible for the outbursts. 
When combining all 13 observations as a function of 
time since the last outburst (albeit not the same outbursts), the quiescent flux 
appears to drop by a factor of 2 over 6.5 years,
or $\Delta F_\mathrm{X} \sim$ $10^{-11}$ ergs cm$^{-2}$ s$^{-1}$ with small ($\times 2$) flares or possible 
orbital variations present. Based upon very limited statistics,
the line-of-sight column density decreases by a factor of a few over the combined 6.5 years of quiescence,
while the power law remained relatively constant. 
The quiescent flux and subsequent, correlated line of sight column depth could be due to a draining of material built-up
at the corotation radius or due to spherical accretion from an isotropic, but diminishing, component of the stellar wind.
Further characterization of the quiescent mode of A\,0535$+$26 will require
more observations over a range of times since the last outburst in order to see if the slow downward trend in the flux 
continues, and to see if one could detect the orbital variations in flux expected for wind accretion in the eccentric orbit. 

\begin{acknowledgments}

We acknowledge the assistance and mentoring of Eric Michelsen concerning Lomb-Scargle analysis and its interpretation,
along with use of his program.Thanks go out to Ron Remillard for supplying the weekly A\,0535$+$26 ASM data for 
the duration of the \textsl{RXTE} mission. 
These results were provided by the ASM/RXTE teams at MIT and at the RXTE SOF and GOF at NASA's GSFC. 
We thank Evan Smith for his excellent efforts in scheduling of the RXTE observations. We thank ISSI 
for hosting and financing international group meetings on modeling and observations of cyclotron lines in High Mass X-ray 
binaries. We acknowledge the GSFC HEASARC for providing extracted spectra and response files for the \textsl{EXOSAT}
observations that allowed for the re-analysis of the A0535$+$26 observations. This work was supported by the 
Bundesministerium f\"ur Wirtschaft und Technologie through DLR grant 50 OR 1113. 
We would like to thank J. E. Davis for the slxfig module which was used to create some of the plots through out this paper.
ACA  thanks the support from  grants AYA2012-39303, SGR2009- 811,  and iLINK2011-0303.

\end{acknowledgments}


\end{document}